\documentclass[
aps,%
11pt,%
final,%
notitlepage,%
oneside,%
onecolumn,%
nobibnotes,%
nofootinbib,
superscriptaddress,%
noshowpacs,%
centertags]%
{revtex4}
\RequirePackage{graphicx}

\def\refitem#1{\relax}

\begin{document}
\title{A novel strong coupling expansion of the QCD Hamiltonian\vspace*{-5mm}}

\author{\firstname{H.-P.} \surname{Pavel}}
\email{hans-peter.pavel@physik.tu-darmstadt.de}
\affiliation{Institut f\"ur Kernphysik, Technische Universit\"at Darmstadt,
D-64289 Darmstadt,
Germany}
\affiliation{Bogoliubov  Laboratory of Theoretical Physics, JINR Dubna,
141980 Dubna, Russia\\[5mm]}

\begin{abstract}
Introducing an infinite spatial lattice with box length a,
a systematic expansion of the physical QCD Hamiltonian in $\lambda = {g}^{-2/3}$ can be obtained.
The free part is the sum of the Hamiltonians of the quantum mechanics of spatially constant fields
for each box, and the interaction terms proportional to $\lambda^n$ with contain n discretised spatial
derivatives connecting different boxes. As an example, the energy of the vacuum and the lowest
scalar glueball is calculated up to order $\lambda^2$ for the case of $SU(2)$ Yang-Mills theory.
\end{abstract}

\maketitle

\section{Introduction}

The quantum Hamiltonian of $SU(2)$ Yang-Mills theory, to which we limit ourselves here for simplicity,
can be obtained \cite{Christ and Lee} by exploiting the time-dependence of the gauge transformations to put
 $A_{a0}(x) = 0~,(a=1,2,3)$, and quantizing the spatial fields in the Schr\"odinger representation,
$ \Pi_{ai}({\mathbf x})=-E_{ai}({\mathbf x})=-i\delta/\delta A_{ai}({\mathbf x})$.
The physical states $\Psi$ have to satisfy the Schr\"odinger equ. and the three non-Abelian Gauss law constraints
\begin{eqnarray}
\label{Hcon}
&& H\Psi = E\Psi~,\quad\quad H = \int d^3{\mathbf x} {1\over 2}
\sum_{a,i} \left[\left(\frac{\delta}{\delta A_{ai}({\mathbf x})}\right)^2+B_{ai}^2(A({\mathbf x}))\right]
\\
\label{Gcon}
&&
G_a({\mathbf x})\Psi = 0~,\quad\quad G_a ({\mathbf x}) = -i\left(\delta_{ac}\partial_i +g\epsilon_{abc}A_{bi}({\mathbf x})\right)
             \frac{\delta}{\delta A_{ci}({\mathbf x})}
\end{eqnarray}
with the chromo-magnetic fields $B_{ai}(A) =\epsilon_{ijk}\left(\partial_j A_{ak}+{1\over 2}g
\epsilon_{abc}A_{bj} A_{ck}\right)$ and the
 generators $G_a({\mathbf x})$ of the residual time-independent gauge transformations, satisfying
$[G_a({\mathbf x}),H]=0~,$ and
$[G_a({\mathbf x}),G_b({\mathbf y})]=ig\delta^3({\mathbf x}-{\mathbf y})\epsilon_{abc}G_c({\mathbf x})~,$
The matrix elements have Cartesian measure
\begin{eqnarray}
\quad \langle\Phi_1|{\cal O} |\Phi_2\rangle=
\int \prod_{{\mathbf x}}\prod_{ik} dA_{ik}({\mathbf x}) \ \Phi_1^* {\cal O}\Phi_2~.
\end{eqnarray}
In order to implement the Gauss laws (\ref{Gcon}) into (\ref{Hcon}), to obtain the physical Hamiltonian,
it is very useful to Abelianise them by a suitable point transformation of the gauge fields.

\section{The physical Hamiltonian of SU(2) Yang-Mills theory and its strong coupling expansion
}

Point transformation to the new set of adapted coordinates \cite{KP1},
the 3 $ q_j\ \ (j=1,2,3)$  and the 6 elements
$S_{ik}= S_{ki}\ \ (i,k=1,2,3)$ of the positive definite
symmetric $3\times 3$ matrix $S$,
\begin{eqnarray}
\label{SymGau}
A_{ai} \left(q, S \right) =
O_{ak}\left(q\right) S_{ki}
- {1\over 2g}\epsilon_{abc} \left( O\left(q\right)
\partial_i O^T\left(q\right)\right)_{bc}\,,
\end{eqnarray}
where \( O(q) \) is an orthogonal $3\times 3$ matrix
parametrised by the $q_i$, leads to an Abelianisation of the Gauss law constraints
\begin{eqnarray}
 G_a\Phi=0\quad
\Leftrightarrow\quad\frac{\delta }{\delta q_i}\Phi=0\quad (\rm{Abelianisation})\nonumber
\end{eqnarray}
Equ. (\ref{SymGau}) corresponds to the symmetric gauge
$\chi_i(A)=\epsilon_{ijk}A_{jk}=0$ .
It has been proven in \cite{KMPR} that, at least for strong coupling, the symmetric gauge exists,
i.e. any time-independent gauge field
can be carried over uniquely into the symmetric gauge, and in \cite{KP1} that both indices of the tensor field $S$
are spatial indices, leaving $S$ a colorless local field.

According to the general scheme of \cite{Christ and Lee},
the correctly ordered physical quantum Hamiltonian  in the symmetric gauge in terms of the colorless
physical variables $S_{ik}({\mathbf{x}})$  and the corresponding canonically
conjugate momenta $P_{ik}({\mathbf{x}})\equiv -i\delta /\delta S_{ik}({\mathbf{x}})$  reads \cite{pavel2}
\begin{eqnarray}
H(S,P)&=& {1\over 2}{\cal J}^{-1}\!\!\int d^3{\mathbf{x}}\ P_{ai}\ {\cal J}P_{ai}
     +{1\over 2}\int d^3{\mathbf{x}}\left(B_{ai}(S)\right)^2\nonumber\\
      & &\!\! -{\cal J}^{-1}\!\!\int\! d^3{\mathbf{x}}\int\!\! d^3{\mathbf{y}}
      \Big\{\Big(D_i(S)_{ma}P_{im}\Big)({\mathbf{x}}){\cal J}
            \langle {\mathbf{x}}\ a|^{\ast}\! D^{-2}(S)|{\mathbf{y}}\ b\rangle
             \Big(D_j(S)_{bn}P_{nj}\Big)({\mathbf{y}})\Big\}
\end{eqnarray}
with the covariant derivative
$~D_i(S)_{kl}\equiv\delta_{kl}\partial_i-g\epsilon_{klm}S_{mi}~$,
the Faddeev-Popov (FP) operator
\begin{equation}
^{\ast}\! D_{kl}(S)\equiv \epsilon_{kmi}D_i(S)_{ml}
   =\epsilon_{kli}\partial_i-g\gamma_{kl}(S)~,
   \quad\quad \gamma_{kl}(S)\equiv S_{kl}-\delta_{kl} {\rm tr} S
\end{equation}
and the Jacobian
$~{\cal J}\equiv\det |^{\ast}\! D|~ $.
The matrix element of a physical operator O is given by
\begin{equation}
\langle \Psi'| O|\Psi\rangle\
\propto
\int \prod_{\mathbf{x}}\Big[dS({\mathbf{x}})\Big]
{\cal J}\Psi'^*[S] O\Psi[S]~.
\end{equation}

The inverse of the FP operator
can be expanded in the number of spatial derivatives
\begin{eqnarray}
\langle {\mathbf{x}}\ k|^{\ast}\! D^{-1}(S)|{\mathbf{y}}\ l\rangle
&=&
-{1\over g}\gamma^{-1}_{k l}({\mathbf{x}})\delta({\mathbf{x}}-{\mathbf{y}})
+{1 \over g^2}\gamma^{-1}_{ka}({\mathbf{x}})\epsilon_{abc}
     \partial_c^{({\mathbf{x}})}
     \left[\gamma^{-1}_{bl}({\mathbf{x}})\delta({\mathbf{x}}-{\mathbf{y}})\right]
\nonumber\\
&&
-{1 \over g^3}\gamma^{-1}_{ka}({\mathbf{x}})\epsilon_{abc}
     \partial_c^{({\mathbf{x}})}\!\!\left[\gamma^{-1}_{bi}({\mathbf{x}})\epsilon_{ijk}
     \partial_k^{({\mathbf{x}})}\!\!
     \left[\gamma^{-1}_{jl}({\mathbf{x}})\delta({\mathbf{x}}-{\mathbf{y}})\right]\right]
+...~.
\label{derexp}
\end{eqnarray}
In order to perform a consistent expansion , also the non-locality in the Jacobian ${\cal J}$
has to be taken into account \cite{pavel2}. The Jacobian ${\cal J}$ factorizes
$
{\cal J}={\cal J}_0 \widetilde{\cal J}
$
with the local
\begin{equation}
{\cal J}_0 \equiv\det |\gamma|=\prod_{\mathbf{x}}\prod_{i<j}
\left(\phi_i({\mathbf{x}})+\phi_j({\mathbf{x}})\right)~,\quad\quad\quad (\phi_i= {\rm eigenvalues}\ {\rm of}\ S)
\end{equation}
and the non-local $\widetilde{\cal J}$, which
can be included into the wave functional
$
\widetilde{\Psi}(S):=\widetilde{\cal J}^{-1/2}\Psi(S)
$
leading to the corresponding transformed Hamiltonian
\begin{equation}
\widetilde{H}(S,P):=\widetilde{\cal J}^{1/2} H(S,P) \widetilde{\cal J}^{-1/2}
=H(S,P)\Big|_{J\rightarrow J_0}+V_{\rm measure}(S)~.
\end{equation}
It is Hermitean with respect to the local measure ${\cal J}_0$ on the cost of extra
terms $V_{\rm measure}$, and can be expanded in the number of spatial derivatives using (\ref{derexp}).

Next, an ultraviolet cutoff $a$ is put by introducing an infinite spatial lattice of granulas
$G({\mathbf{n}},a)$, here cubes of length $a$,
situated at sites ${\mathbf{x}}=a {\mathbf{n}}$  $({\mathbf{n}}\in Z^3)$, and
considering the averaged variables
\begin{equation}
\label{average}
S({\mathbf{n}}) :=  \frac{1}{a^3}\int_{G({\mathbf{n}},a)} d{\mathbf{x}}\ S({\mathbf{x}})
\nonumber
\end{equation}
and discretised spatial derivatives relating the $S({\mathbf{n}})$ of different granulas
(see \cite{pavel2} for details).

After an appropriate rescaling of the dynamical fields
a novel strong coupling expansion of the Hamiltonian in $\lambda=g^{-2/3}$ can be obtained \cite{pavel2}
\begin{equation}
\label{Hexpan}
\widetilde{H} =\frac{g^{2/3}}{a}\left[{\cal H}_0+\lambda \sum_\alpha {\cal V}^{(\partial)}_\alpha
                             +\lambda^2 \left(\sum_\beta {\cal V}^{(\Delta)}_\beta
                             +\sum_\gamma{\cal V}^{(\partial\partial\neq\Delta)}_\gamma\right)
                             + {\mathcal{O}}(\lambda^3)\right]~.
\label{Hexpansion}
\end{equation}
as an alternative to existing strong coupling expansions \cite{Kogut} and \cite{Muenster} based on Wilsonian
lattice QCD.
The "free part" in (\ref{Hexpansion}) is just the sum of Hamiltonians
${\cal H}_0 =\sum_{\mathbf{n}}{\cal H}^{QM}_0({\mathbf{n}})$
of Yang-Mills quantum mechanics of spatially constant fields  \cite{Luescher}-\cite{pavel} at each site,
and the ${\cal V}^{(\partial)}_\alpha$ and ${\cal V}^{(\Delta)}_\beta$ are interaction parts,
relating different sites. The local measure ${\cal J}_0=\prod_{\mathbf{n}}{\cal J}^{QM}_0({\mathbf{n}})$
is correspondingly the product of the quantum mechanical measures at each site.
In terms of the principal-axes variables of the positive definite symmetric $3\times 3$ matrix field $S$
\begin{eqnarray}
S  =  R^{T}(\alpha,\beta,\gamma)\ \mbox{diag}\ ( \phi_1 , \phi_2 , \phi_3 ) \
R(\alpha,\beta,\gamma)~,
\end{eqnarray}
with the \( SO(3)\) matrix  \({R}\)
parametrized by the three Euler angles \(\chi=(\alpha,\beta,\gamma )\), we find
\begin{eqnarray}
\label{range}
{\cal J}_0^{QM} \rightarrow  \sin\beta
\prod_{i<j}\left(\phi_i^2-\phi_j^2\right)
\quad
\rightarrow \quad 0<\phi_1<\phi_2<\phi_3\quad (\rm{principle}\ \rm{orbits})~.
\end{eqnarray}
and (with the intrinsic spin angular momenta $\xi_i$ )
\begin{eqnarray}
\label{calH0}
{\cal H}^{QM}_0 = {1\over 2}\sum^{\rm cyclic}_{ijk}
\Big[ \pi_i^2
-{2i\over \phi_j^2-\phi_k^2}
\left(\phi_j\pi_j
-\phi_k \pi_k\right)
+ \xi^2_i {\phi_j^2+\phi_k^2
\over (\phi_j^2-\phi_k^2)^2}
 +\phi_j^2 \phi_k^2
\Big]~.
\end{eqnarray}
Its low energy spectrum and eigenstates   at any site $\mathbf{n}$
\begin{equation}
{\cal H}_0^{QM}({\mathbf{n}})  |\Phi_{i,M}^{(S)\pm}\rangle_{\mathbf{n}} =
\epsilon^{(S)\pm}_i ({\mathbf{n}}) |\Phi_{i,M}^{(S)\pm}\rangle_{\mathbf{n}}~,
\end{equation}
characterised by the quantum numbers of spin $S,M$ and parity P, are known
with high accuracy \cite{pavel}.
Hence the eigenstates of ${\cal H}_0$ in (\ref{Hexpansion}) are free glueball excitations of the lattice.
The interactions ${\cal V}$ (\ref{Hexpan}) can be included using perturbation theory in $\lambda$.

\section{Calculation of the glueball spectrum up to order $\lambda^2$}

Using 1st and 2nd order perturbation theory in $\lambda$ give the results \cite{pavel2}
\begin{eqnarray}
E_{\rm vac}^{+}={\cal N}\frac{g^{2/3}}{a}\Bigg[4.1167+29.894\lambda^2+{\cal O}(\lambda^3)\Bigg]~,
\end{eqnarray}
for the energy of the interacting glueball vacuum and
\begin{eqnarray}
 E_1^{(0)+}(k)-E_{\rm vac}^{+}  &=&
 \left[\ 2.270 + 13.511\lambda^2 + {\cal O}(\lambda^3)\right]\frac{g^{2/3}}{a}
 +0.488 \frac{a}{g^{2/3}} k^2 +{\cal O}((a^2 k^2)^2) ~,
\end{eqnarray}
for the energy spectrum of the interacting spin-0 glueball, up to $\lambda^2$
for the $(+)$ b.c. and similar results for the $(-)$ b.c.
The first, zeroth order numbers, correspond to the result of Yang-Mills quantum mechanics.
Note that Lorentz invariance asks for energy momentum relation $ E=\sqrt{M^2+k^2}\simeq M + (2 M)^{-1}\ k^2$.
The result, shown here for the scalar glueball, which limits itself to the
terms in the Hamiltonian containing the Laplace-operator $\Delta$ as a first step,
violates this condition by about a factor of two.
Including all spin-orbit coupling terms in the Hamiltonian dropped in this first
approach and considering all possible $J=L+S=0$ states is expected restore Lorentz invariance.

To study the coupling constant renormalisation in the IR,
consider the physical glueball mass
\begin{equation}
\label{flow}
M = \frac{g_0^{2/3}}{a}\left[\mu +c g_0^{-4/3}\right]~.
\end{equation}
Independence of the box size $a$ is given for the two cases,
$g_0 = 0 $ or $g_0^{4/3}=-c/\mu$.
The first solution corresponds to the perturbative fixed point, and the second,
if it exists $(c<0)$, to an infrared fixed point.
My result for the lowest spin-0 glueball $ c_1^{(0)}/\mu_1^{(0)}=5.95   $
suggests, that no infrared fixed points exist,
in accordance with the corresponding result of Wilsonian lattice QCD \cite{Creutz}.
Solving the above equation (\ref{flow}) for positive $(c>0)$ one obtains
\begin{eqnarray}
g_0^{2/3}(M a) =\frac{M a}{2\mu}
     +\sqrt{\left(\frac{M a}{2\mu}\right)^2
     -\frac{c}{\mu}}~,\quad
     \  a >  a_{c} :=  2\sqrt{c \mu}/M
\label{largebox}
\end{eqnarray}
with the physical glueball mass $M$.
For a typical \cite{Teper} $ M \sim 1.6\ {\rm GeV}$ we find $a_{c} \sim 1.4\ {\rm fm}$.
Comparing the behaviour of the
bare coupling constant (\ref{largebox}),
obtained for boxes of large size a, with those obtained for small boxes in \cite{Luescher,Luescher and Muenster},
should lead to information about the intermediate region, including the possibility
of the existence of phase transitions.

The generalisation of the approach to include quarks is possible \cite{pavel3} and under current investigation.

\begin{acknowledgments}
I would like to thank A. Dorokhov and  J. Wambach for their interest.
Financial support by the LOEWE-Program HIC for FAIR is gratefully acknowledged.
\end{acknowledgments}


\end{document}